# Quantification of Dislocation-Precipitate Interactions

Amirreza Keyhani[1] and Reza Roumina[2]

[1)] *The George W. Woodruff School of Mechanical Engineering, School of Materials Science and Engineering, Georgia Institute of Technology, Atlanta, GA 30332-0405, USA*
[2)] *School of Metallurgical and Materials Engineering, College of Engineering, University of Tehran, Tehran, Iran*

**Abstract**

The present research is the first attempt to systematically quantify the dislocation-precipitate interaction in terms of applied shear stress, precipitate resistance, and the required time to reach the critical state of dislocation-precipitate interaction when a dislocation line is about to pass through precipitates. To model the dislocation-precipitate interaction, we adopt a modified three-dimensional dislocation dynamics. Using the present modeling approach, which employs three-dimensional dislocation dynamics simulations, we obtain thousands of data points, accounting for various precipitate resistances, applied shear stresses, and precipitate spacing. The material of reference is Copper (Cu). From the simulations, which quantify the dislocation-precipitate interaction in terms of the applied shear stress, precipitate resistance scale, and dislocation-precipitate interaction time, we found a universal equation. The dislocation-precipitate interaction time versus precipitate resistance and stress, referred to as the "dislocation-precipitate interaction map," determines the "pass" or "no-pass" state of the interaction. Using this map, we incorporate the dislocation-precipitate interaction time in a two-dimensional multiscale framework which adopts the dislocation dynamics approach at the micro-scale and the finite element method at the macro-scale. We use this framework to model the mechanical behavior of free-standing copper thin films. The results show a dual effect of the dislocation-precipitate interaction time on the hardening level.

***Keywords:*** Dislocation-precipitate interaction time, line dislocation dynamics, multiscale model, precipitates resistance scale, thin film

## 1. Introduction

Researchers have used precipitation hardening to design alloys with the desired ductility and strength; however, this phenomenon was not well understood until the 20[th] century as the development of transmission electron microscopy enabled us to observe dislocations. Precipitates restrict the motion of dislocations, resulting in higher strength as the movement of each dislocation contributes to the total macroscopic plastic deformation. To provide a more comprehensive understanding of the precipitation hardening phenomenon, the physics of dislocation motions and their interactions with obstacles should be quantified more effectively. To explore the dislocation-precipitate interactions, researchers have widely adopted computational approaches. Early computational attempts basically focused on the movement of a dislocation line through a random array of precipitates and the geometrical deformation of the dislocation line [1, 2]. Generally, precipitates are considered dimensionless and modeled as obstacles against dislocation motions. Geometrical approaches successfully capture critical resolved shear stress (CRSS). For more details on the development and limitations of the geometrical approaches, see a review by Ardell [3].

---

[1] akeyhani3@gatech.edu
[2] roumina@ut.ac.ir

The development of the dislocation dynamics approach (DD) [4-7], a computational methodology for modeling dislocation motions and their interactions at the micron scale, and the coupling of DD with continuum modeling such as FEM [8-12], BEM [13], and recently XFEM [14] provide the capability of solving more complex physical and realistic models. Several studies have addressed physical issues of dislocation-precipitate interactions by applying dislocation dynamics such as precipitate shearing by a dislocation [15], the role of the matrix-precipitate shear modulus difference [13], and misfit stresses at the boundary of a precipitate [16]. In addition to applying early geometrical approaches and dislocation dynamics, studies have also adopted molecular dynamics (MD) [17, 18] and multiscale approaches [19-22] to examine dislocation-precipitate interactions.

The dislocation-precipitate interaction results from several factors: (1) the matrix-precipitate shear modulus difference, (2) misfit strains caused by thermal effects, (3) misfit dislocations at the boundary of precipitates, the result of matrix/precipitate different crystalline structures, (4) a change in core energy as the dislocation passes through precipitates, etc [23]. As micro-scale computational approaches for modeling dislocation-precipitate interactions are not capable of treating all mentioned factors, most micro-scale studies are limited to the first factor [10, 13, 15, 24-29]. Monnet defined precipitates as a friction stress opposed to the dislocation movement [30]. Mohles conducted pioneer research to model lattice mismatches [31-33] and imperfect matrix-precipitate interfaces [34]. Duesbery and Sadananda modeled the passage of edge and screw dislocations through an array of spherical coherent obstacles for several obstacle sizes, array dimensions, and misfit values, using computer techniques adapted from molecular dynamic methods [35]. Keyhani et al. [36] developed an efficient computational technique that implicitly accounts for all mentioned factors by introducing the precipitate resistance scale. In addition to the above-mentioned factors, the geometrical arrangement of precipitates plays a role and adds further complexity to the nature of the dislocation-precipitate interaction. Because of the complexity of this interaction and the diversity of determining factors, most studies focus just on the role of a few factors, resulting in a limited understanding of the interaction. Furthermore, they have not addressed the relationship between the precipitate resistance and the required time it takes to reach the critical state of the dislocation-precipitate interaction.

A key parameter analyzed in this research is the required time it takes a dislocation to reach the critical state of the dislocation-precipitate interaction, or the time difference between when a dislocation line initially starts to deform under the influence of precipitates and when the dislocation line is about to pass through the precipitates, referred to here as the "critical time." As the applied shear stress level increases, the dislocation-precipitate interaction undergoes a transition, which may affect the interaction time. At low stress levels, dislocations could bypass precipitates resulting from thermal fluctuations at an attempt frequency governed by lattice dynamics; however, at high stress levels relative to the critical resolved shear stress (CRSS), the overdriven dynamics of DD dominates the thermally-assisted bypass. As a result, the critical time at which a dislocation-precipitate interaction occurs depends mostly on the applied shear stress and precipitate properties. In the current study, simulations are based on the internal energy barrier (enthalphic part) associated with dislocation fields in DD. The study, however, does not account for the entropic barrier associated with anharmonic effects in the thermally activated process controlling the critical time either at low stress levels (creep conditions) or high homologous temperatures [37]. The temperature dependence of obstacle-dislocation interactions in solid solution hardening alloys are discussed in a review article by Kocks [38]. Another paper analyzed the strain rate sensitivity (SRS) in terms of obstacle strength and distributions, temperature, and applied shear stress [39, 40]. For further details about the dislocation-precipitate interaction see [23].

The aim of the present research is to open a new window to understanding the dislocation-precipitate interaction. In order to accomplish this, this study performs extensive computational simulations that find one of the most fundamental relationships for the dislocation-precipitate interaction in terms of applied shear stress, precipitate resistance, and the critical time, referred to as the "dislocation-precipitate interaction map." We construct the dislocation-precipitate interaction map for copper (Cu) with various precipitate resistances, and spacing.

## 2. Framework of the analysis
*2.1. Computational scheme*

To model the interaction between dislocation lines and arrays of equally spaced precipitates, Keyhani et al. [32] introduced a recent methodology based on the standard line dislocation dynamics (DD) simulation code, DDLab. We use the DD procedure to discretize a dislocation line into straight segments defined by two end nodes; the dislocation drag governs mobility relating the nodal forces to the nodal velocities [43]. The usual overdriven dynamics of DD assumes that the drag coefficient is a constant. Since the dislocation-phonon interactions, which lead to a drag force, vary with short-range dislocation-precipitate interaction, this approach is an approximation. In addition, we assume that the simulations are based on elastic isotropy.

In the adopted approach, when a dislocation line encounters a precipitate, the corresponding nodes of the dislocation closer than a specific distance to the precipitate are locked. Hence, the free part of the dislocation line bends, so the related shear stress, which the dislocation exerts on the precipitate, increases. If the resulting shear stress from the dislocation curvature reaches resistance by the precipitate, the locked nodes unlock; therefore, the dislocation begins to pass through the precipitate through shearing. If the resulting shear stress from bending the dislocation line does not reach the precipitate resistance, the dislocation line forms a loop to pass by the precipitate.

The precipitate resistance scale, $R$, is originally introduced and quantified in the authors' former research [36],

$$R = \frac{\tau_p}{\tau_{max}} \tag{1}$$

where $\tau_p$ is the shear stress against dislocation movement due to the presence of precipitates and $\tau_{max} = \mu b / D_1$ is the maximum possible shear stress exerted to the precipitate by a dislocation when the radius of dislocation curvature is equal to the radius of the first Orowan loop, $D_1$. $\mu$ is the shear modulus of the matrix and $b$ is the magnitude of the Burgers vector. The resistance scale, $R$, is equal to unity for non-shearable precipitates, representing the Orowan mechanism, and less than unity for shearable precipitates. The adopted methodology is not capable of modeling prismatic loops, known as the Hirsch mechanism, which may form in FCC crystals as well as Orowan loops [44, 45].

*3.2. Simulation setup*

The critical time for the interaction of a straight edge dislocation line with an array of collinear equally spaced precipitates with similar resistance and diameter is calculated, Fig. 1. According to the adopted computational methodology, the critical time is the time difference between when the first node of a dislocation line is locked as it reaches to the distance from the precipitate equals to the diameter of the corresponding Orowan loop and when the dislocation line starts to pass the precipitate either by forming Orowan loops or by cutting through the precipitate.

The critical time of the dislocation line interacting with precipitates is calculated for thousands of interactions by varying the applied stress, precipitate resistance, and ratio of $L/D$. The applied shear stress, $\tau_{app}$, is considered to be a fraction of the critical shear stress for non-shearable precipitates ($\tau_c^{R=1}$) [46],

$$\tau_c^{R=1} = \frac{\mu b}{2\pi L} \ln\left(\frac{\bar{D}}{r_0}\right) \tag{2}$$

where $L$ is the internal spacing of the precipitates, $\bar{D} = (D^{-1} + L^{-1})^{-1}$ in which $D$ is the precipitate diameter considered to be 100 nm for all simulations, and $r_0$ is the dislocation core radius. Similar to the Frank-Read mechanism, the time for dislocation-precipitate interaction at the critical state goes to infinity if the applied shear stress is equal to the critical stress, $\tau_c^{R=1}$; therefore, we introduce $\tau_c' = 1.05\tau_c^{R=1}$ and $t_c'$ is the corresponding time at the critical state for $\tau_c'$. We perform simulations for Cu, an FCC crystal, and three $L/D$ ratios within the range of 5 to 10 [13]. Table 1 presents all studied conditions and material properties. The maximum dislocation segment length and the maximum dislocation movement in each step are 10 nm, which are one-tenth of the precipitate diameter. Therefore, there are at least 10 dislocation segments encountering with a precipitate, which can pass through the precipitate at least in ten simulation steps.

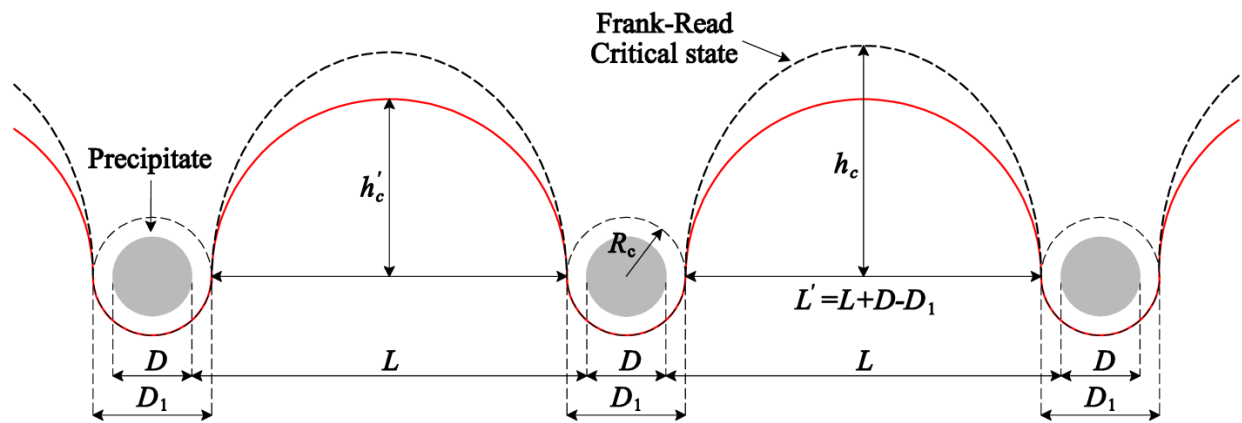

Fig. 1. The interaction geometry: a dislocation line encounters an array of collinear equally spaced precipitates with similar properties. The red solid line and dashed black line present the dislocation at the critical state of dislocation-precipitate interaction and equivalent Frank-Read mechanism with the length of $L'$, respectively.

Table 1. Simulation configurations and material properties.

| | No. of conditions | Used parameters | | | | | |
|---|---|---|---|---|---|---|---|
| $\tau_{app}$ | 6 | $[0.25 \quad 0.50 \quad 0.75 \quad 1.00 \quad 1.50 \quad 2.00]\tau'_c$ | | | | | |
| $L/D$ | 3 | $[5.0 \quad 7.5 \quad 10.0]$ | | | | | |
| Material | Crystal | $\mu$ (GPa) | $\upsilon^*$ | $b$ (nm) | $B_e^{**}$ (Pa.s) | $B_s^{\dagger}$ (Pa.s) |
| | Cu (FCC) | 63.2 | 0.305 | 0.256 | $2.31\times10^{-6}$ | $9.82\times10^{-6}$ |

*Poisson's ratio
**Drag coefficient for an edge dislocation [47]
† Drag coefficient for a screw dislocation [47]

## 3. Results

*3.1. Dislocation-precipitate interaction versus Frank-Read nucleation mechanism*

As depicted in Fig. 1, a dislocation line flanking between two non-shearable precipitates behaves similar to a Frank-Read source with the equivalent length of $L'$. However, the critical time of dislocation-precipitate interaction and the nucleation time of the Frank-Read source are not identical. This is because of the fact that the dislocation-precipitate interaction time is based on the time at the critical state of the dislocation segment stopped by a precipitate while the Frank-Read nucleation time refers to the critical state of the dislocation line segment bounded between two precipitates. The critical radius of the dislocation segment stuck by a precipitate, $R_c$, is equal to the Orowan loop radius. The geometry of the dislocation line at the nucleation state of the Frank-Read mechanism is an oval shape, $2h_c/L' = 1.5$ for an edge dislocation [48]; however, the geometry of the dislocation line pinned between two non-shearable spherical precipitates at the critical state of dislocation-precipitate interaction is semicircular, $2h_c/L' = 1$. In order to overcome an array of shearable precipitates, the dislocation line undergoes less bending as the corresponding CRSS is lower than non-shearable precipitates, which means that the free part of the dislocation line requires to move less in the outward direction of the precipitate array to reach the critical state, $2h_c/L' < 1$.

Figure 2 compares the dislocation-precipitate interaction time for non-shearable precipitates calculated by DD with the nucleation time of the Frank-Read source obtained from the following equation [49],

$$t_{nuc} = \frac{BL'}{\tau_{nuc}b} \frac{\eta}{2} \frac{1}{\zeta} \left\{ 1 + \frac{1}{\zeta} \left[ \frac{1}{2} \ln \frac{2\zeta - 2}{\zeta} + \frac{1}{\sqrt{\zeta^2 - 1}} \left( \tan^{-1} \frac{\zeta - 1}{\sqrt{\zeta^2 - 1}} + \tan^{-1} \frac{1}{\sqrt{\zeta^2 - 1}} \right) \right] \right\} \quad (3)$$

by assuming a circular shape at the critical state for two crystals and three ratios of $L/D$. As illustrated in Fig. 2, there is a good agreement between the time obtained from DD and the one calculated from Eq. (3). The parameter $B$ in Eq. (3) is the drag coefficient, $L'$ is the equivalent length of Frank-Read sources, $\eta = 2h_c/L' = 1$, $\tau_{nuc}$ is the nucleation stress of the Frank-Read source, and $\zeta = \tau/\tau_{nuc}$.

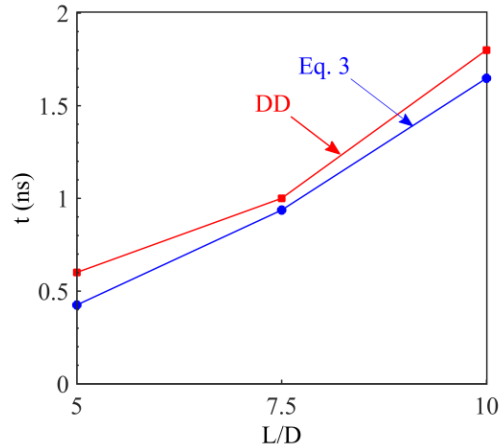

Fig. 2. Comparison of the dislocation-non-shearable precipitates $(R = 1)$ interaction time and the nucleation time of the Frank-Read source with equivalent length and circular shape.

*3.2. Dislocation-precipitate interaction map*

A systematic computational study based on dislocation dynamics approach are performed resulting thousands of data points for the critical time of dislocation-precipitate interaction versus applied shear stress and precipitate resistance. In each simulation, the critical time for the dislocation-precipitate interaction at a given applied shear stress and precipitate resistance is calculated. The snapshots of the dislocation-precipitate interaction with $L/D=5$ under the applied shear stress of $\tau = 1.5\tau'_c$ for an impenetrable precipitate $(R=1)$ and penetrable precipitate $(R=0.5)$ are presented in Figs. 3 and 4, respectively. According to Figs. 3 and 4, the critical time increases for stronger precipitates.

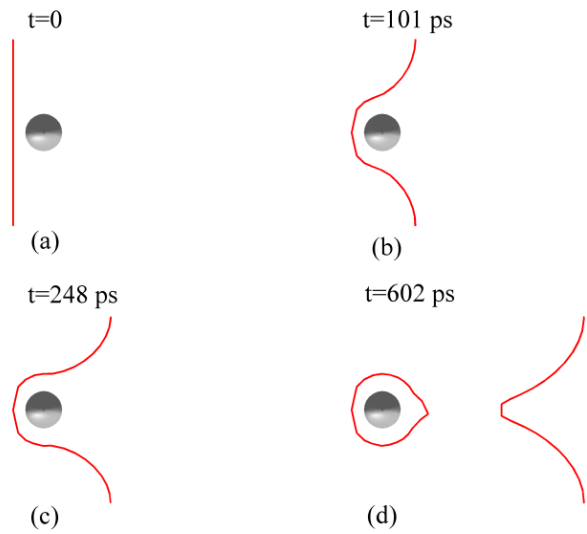

Fig. 3. Snapshots of an edge dislocation interaction with an impenetrable precipitate under $\tau = 1.5\tau'_c$ and $L/D=5$. Figure 3(c) shows the critical state.

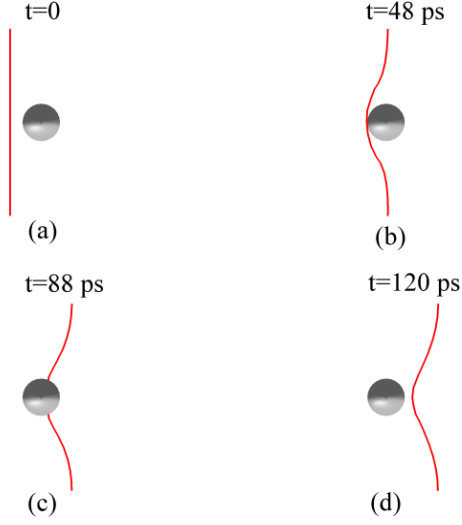

Fig. 4. Snapshots of an edge dislocation interaction with a penetrable precipitate, $R=0.5$, under $\tau =1.5\tau'_c$ and $L/D=5$. Figure 2(b) shows the critical state.

Figure 5 presents the results of all simulations and we cast all the simulation results in one equation with the following form,

$$\left(\frac{R}{a_1}\right)^{s_1} + \left(\frac{t^*}{a_2}-1\right)^{s_2} = 1 \qquad (4)$$

where $\tau^* = \tau_c^R/\tau'_c$, $t^* = t_c^R/t'_c$, $\tau_c^R$ and $t_c^R$ are the critical stress and time for the precipitate resistance $R$, respectively. We pick the form of Eq. (4) based on the results trend and calculate the constants $s_1$ and $s_2$ by the least-squares regression method. Table 2 presents values of the parameters in Eq. (4). Eq. (4) represents a supper-ellipse function, which is the most general relation for quantification of an edge dislocation interaction with an array of collinear equally spaced precipitates. The physical domain for this function is $0 \le R \le 1$ and $0 \le t^* \le 1$, shown by a rectangular with a solid line in Fig. 6.

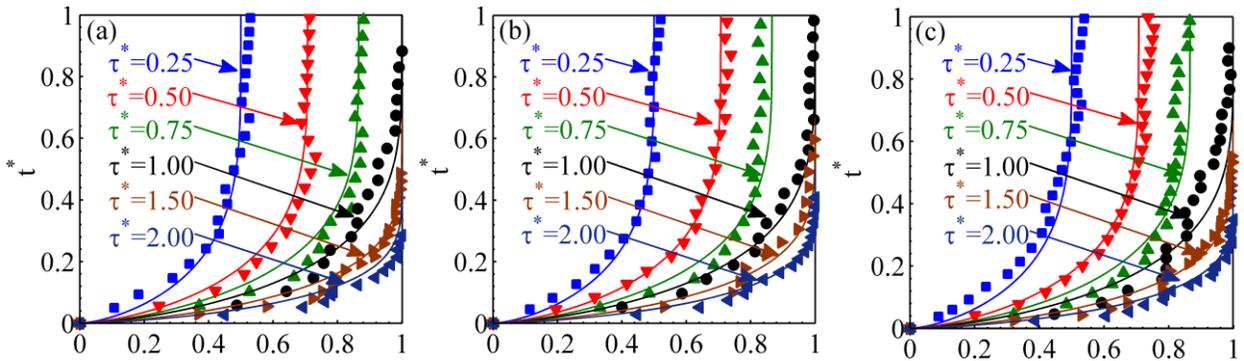

Fig. 5. $t^*$ versus $R$ and $\tau^*$ for several cases. Cu crystal: (a) $L/D = 5$, (b) $L/D = 7.5$, and (c) $L/D = 10$.

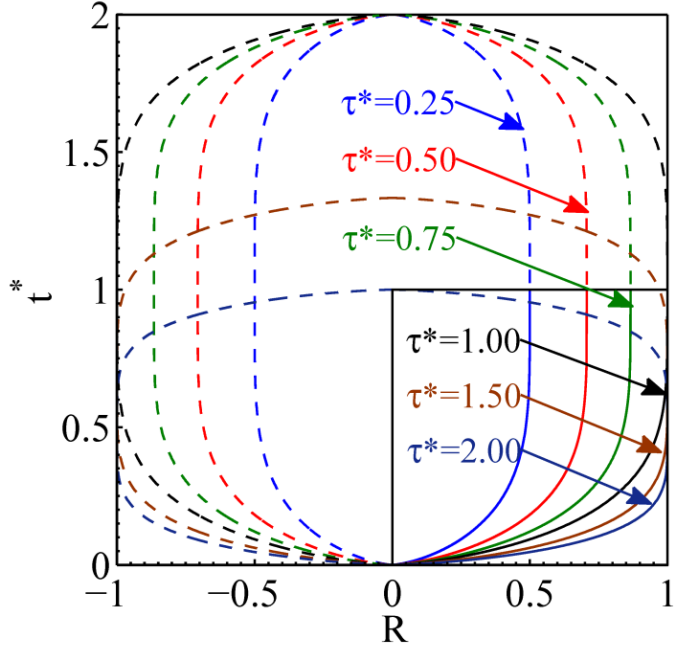

Fig. 6. Dislocation-precipitate interaction map: The rectangular solid line presents the physical domain of the map.

Table 2. Values of the parameters in Eq. (4).

|  | $a_1$ | $a_2$ | $s_1$ | $s_2$ |
|---|---|---|---|---|
| $\tau^* \leq 1$ | $\sqrt{\tau^*}$ | 1 | $\dfrac{3}{2}$ | $\dfrac{9}{2}$ |
| $\tau^* > 1$ | 1 | $1/\tau^*$ | | |

The normalization of the applied shear stress and time at critical state by the critical shear stress and time for $\tau'_c = 1.05 \tau_c^{R=1}$ leads to an independent relation of geometry, $L/D$, and material properties in Eq. (4). Each data point on the dislocation-precipitate interaction map corresponds to a specific interaction case with the known applied shear stress and the precipitate resistance scale. The third parameter, the time at the critical state, determines the state of the interaction. There are three possibilities for the dislocation-precipitate interaction, which decreases to the two cases when $\tau^* > 1$. Figure 7 illustrates the three domains corresponding to each dislocation-precipitate interaction state. Inside the super-ellipse, labeled by 1 in Fig. 7, represents the interaction cases that the dislocation has reached to the critical state. The vertical difference between a point, for instance point A, and the super-ellipse boundary shows the time after the dislocation reached to the critical state, $t^+$. Every point outside of the super-ellipse refers to the dislocation-precipitate interaction state in which the dislocation does not reach the critical state. However, for the resistance in the range of $0 \leq R \leq R(t^* =1, \tau^*)$, point B in domain 2, the dislocation will reach to the critical state after the time $t^-$. Domain 3, outside the super-ellipse and $R(t^* =1, \tau^*) < R \leq 1$, corresponds to the state of the interaction that the dislocation does not reach the critical state as the applied shear stress is not large enough. Domain 3 vanishes when $\tau^* > 1$ since the applied shear stress is large enough to push the dislocation to the critical state.

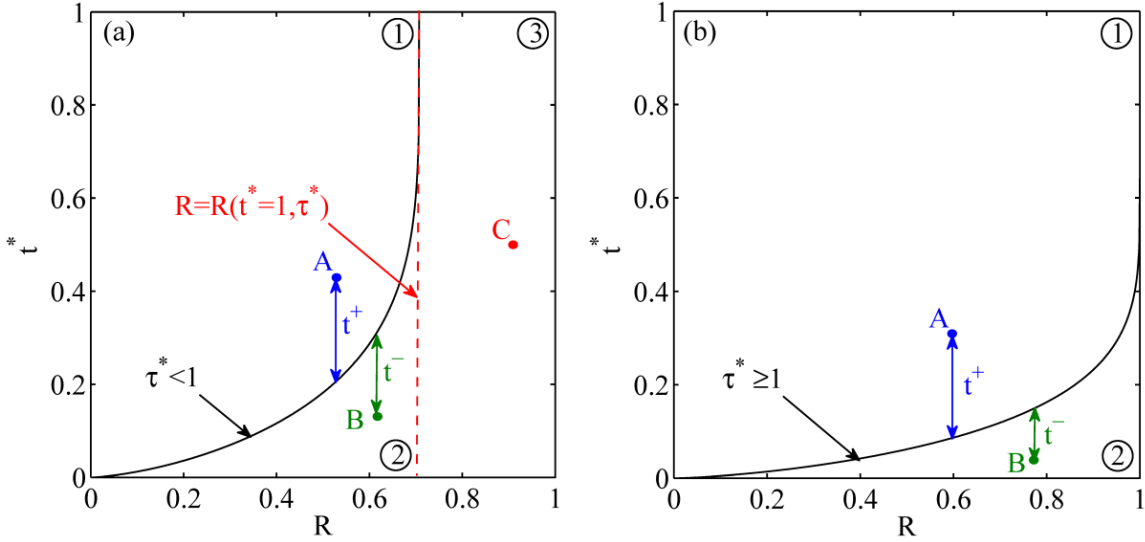

Fig. 7. Representation of different regimes for the dislocation-precipitate interaction; (a) $\tau^* \leq 1$, (b) $\tau^* > 1$.

*3.3. Dislocation-precipitate interaction time under the time-dependent applied shear stress*

Equation (4) is obtained based on the constant applied shear stress but it can be extended for the time-dependent applied shear stress. The tangent to the dislocation-precipitate interaction map at each point represents the time increment due to the increment in $R$ for the corresponding applied shear stress of $\tau^*$. Therefore, the dislocation-precipitate interaction map can be stated in the following discretized form,

$$R_p = \sum_{n=1}^{m} \Delta R^n \qquad (5)$$

where $R_p$ is the resistance scale of precipitate arrays. The time increment, $\Delta t^{*n}$, corresponding to $\Delta R^n$ can be calculated by the following equation, Fig. 8,

$$\Delta t^{*n} = \left. \frac{\partial t^*}{\partial R} \right|_{\substack{\tau^* = \tau^{*n-1} \\ R = R^{n-1}}} \Delta R^n \qquad (5)$$

here, $\partial t^*/\partial R$ is

$$\begin{cases} \dfrac{\partial t^*}{\partial R} = \dfrac{R^{1/2}}{3\tau^{*3/4}\left(1+R^{3/2}\tau^{*-3/4}\right)^{7/9}} & \tau^* \leq 1 \\ \dfrac{\partial t^*}{\partial R} = \dfrac{R^{1/2}}{3\tau^*\left(1+R^{3/2}\right)^{7/9}} & \tau^* > 1 \end{cases}, \qquad (6)$$

$\tau^{*n-1}$ and $R^{n-1}$ are the applied shear stress and the resistance scale at the step $n-1$. By summing all time increments the time at the critical state of dislocation-precipitate interaction can be obtained,

$$t^* = \sum_{n=1}^{m} \Delta t^{*n} \qquad (7)$$

The time at the critical stage where a dislocation line is on the verge of passing a precipitate for several time-dependent applied shear stresses is presented in Fig. 9. According to Fig. 9, the results from DD simulation and the above-mentioned approximation are in a good agreement.

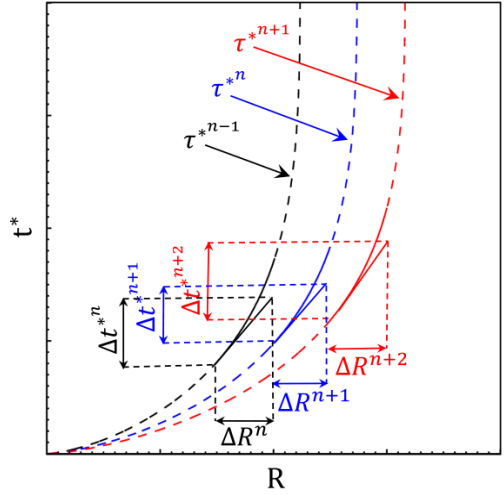

Fig. 8. Discretization of the dislocation-precipitate interaction map.

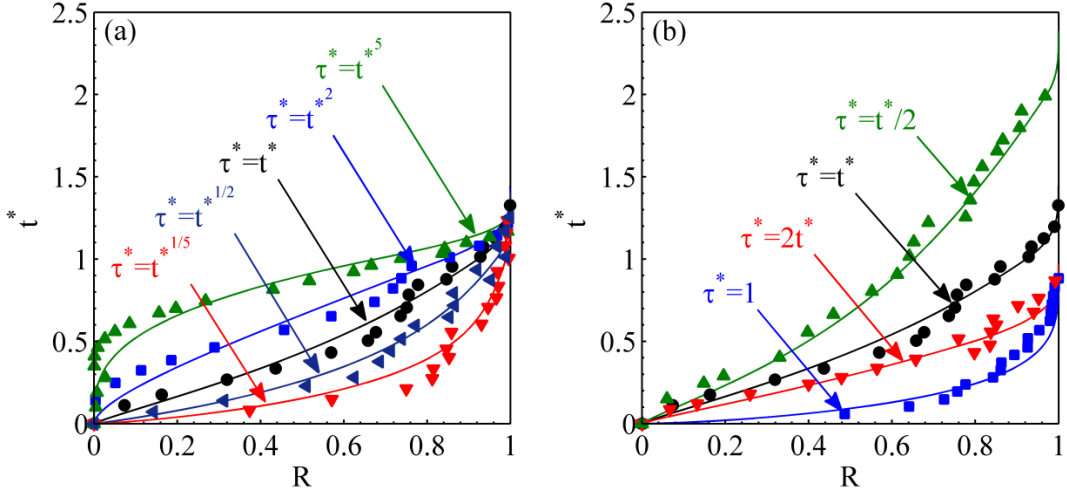

Fig. 9. Time at the critical state for different time-dependent applied shear stresses: (a) $\tau^*$ is a power function of $t^*$, (b) $\tau^*$ varies linearly with $t^*$. The data points represent the DD simulation results and solid lines show the present approximation.

*3.4. Multiscale modeling of plasticity in a precipitate hardened copper thin film*

Hardening mechanisms at micro-scale plastic deformation are fundamentally in three dimensions. Therefore, some physics are missing in the two-dimensional dislocation dynamics. However, the computational efficiency of two-dimensional dislocation dynamics with respect to three-dimensional approach allows simulating a realistic number of dislocations and grains. This fact motivates a few studies to incorporate three-dimensional mechanisms in two-dimensional dislocation dynamics approach [50-52]. Dislocation-precipitate interaction is an important hardening mechanism which has not been modeled precisely in two-dimensional dislocation dynamics approach. So far, precipitates are modeled as an obstacle with a specific resistance. In this model, a dislocation stops when it encounters an obstacle unless the critical resolved shear stress is higher than the obstacle resistance. To build a more physical dislocation-precipitate interaction model for two-dimensional simulations, we incorporate the extended dislocation-precipitate interaction map, presented in Section 3.3, in a two-dimensional multiscale framework to analyze the mechanical response of a freestanding copper thin film. This framework uses the two-dimensional dislocation dynamics (DD) to analyze dislocation motions and the finite element method (FEM) to correct the stress field due to boundary conditions and capture the deformation. To bridge the gap between DD and FEM, the framework uses the following equation [11, 12],

$$\mathbf{KU} = \mathbf{f}_{ext} + \mathbf{f}_B + \mathbf{f}_\infty + \mathbf{f}_P \tag{8}$$

where $\mathbf{K}$ and $\mathbf{U}$ are the stiffness matrix and the displacement vector, respectively. $\mathbf{f}_{ext}$ is the external force vector. $\mathbf{f}_B$ results from the presence of dislocations and $\mathbf{f}_\infty$ treats boundary conditions. $\mathbf{f_p}$ results from dislocation motions and yields the equivalent plastic strain at the finite element analysis,

$$\begin{aligned} \mathbf{f}_{ext} &= \int_{\Gamma} \mathbf{\bar{t}N} d\Gamma \\ \mathbf{f_B} &= \int_{\Omega} \mathbf{S_D B} d\Omega \\ \mathbf{f}_{\infty} &= -\int_{\Gamma} \mathbf{t}^{\infty} \mathbf{N} d\Gamma \\ \mathbf{f_p} &= \int_{\Omega} \mathbf{D\varepsilon_p B} d\Omega \end{aligned} \quad (8)$$

in which $\mathbf{\bar{t}}$ and $\mathbf{t}^{\infty}$ are the applied traction and the resulting traction from the presence of dislocations on the boundary $\Gamma$, respectively. $\mathbf{S_D}$ is the average stress field due to the presence of dislocations in the finite domain, $\Omega$, which is identical to each element of the finite element analysis. $\mathbf{N}$ is the vector of shape functions, $\mathbf{B} = \nabla \mathbf{N}$. $\mathbf{\varepsilon^p}$ is the plastic strain vector resulting from dislocation motions, and $\mathbf{D}$ is the elastic stiffness tensor.

Dislocation motion is governed by a mobility law which relates the applied stress on a dislocation to its velocity. The linear dependence of velocity on stress is valid only for a limited range of applied stress [53-55]. Here, we use a nonlinear mobility equation [56],

$$v = v_0 \left( \tau / \tau_0 \right)^m \quad (8)$$

where $m$ and $\tau_0$ are material constants and $v_0$ is unit velocity. For copper, $m = 0.7$ and $\tau_0 = 2.7$ KPa [56]. The macroscopic strain $\mathbf{\varepsilon_p}$ results from the movement of each dislocation,

$$\mathbf{\varepsilon_p} = \sum_{i=1}^{N} \frac{l_i v_i}{2 V_{RVE}} \left( \mathbf{n}_i \otimes \mathbf{b}_i + \mathbf{b}_i \otimes \mathbf{n}_i \right) \quad (9)$$

with summation over the number of dislocations $N$. $l_i$, $v_i$, $\mathbf{n}_i$ and $\mathbf{b}_i$ are the dislocation segment length, the magnitude of glide velocity, unit normal vector to the slip plane, and the Burgers vector of dislocation, respectively. $V_{RVE}$ is the volume of the finite element in which the dislocation glide occurs.

We carry out simulations on a freestanding $3 \times 1$ $\mu$m copper thin film with identical material properties used in Section 2. The film is subjected to uniaxial tension with a strain rate of $10^4 s^{-1}$. The model contains three glide planes that differed by 60 degrees as illustrated schematically in Fig. 10. The thin film is discretized by $30 \times 90$ ordinary 4-node cubic finite elements. The free surface condition is applied on the top and bottom surfaces of the film by the term $\mathbf{f}_{\infty}$ in Eq. . The density of Frank-Read sources and initial dislocations are 30 $\mu m^{-2}$ and 10 $\mu m^{-2}$, respectively. We select the strength of the Frank-Read sources randomly from a Gaussian distribution with an average value of $\bar{\tau}_{nuc} = 49.4$ MPa and a standard deviation of $0.2\bar{\tau}_{nuc}$. The nucleation time of Frank-Read sources is $23.1$ ns. The mean nucleation distance $L_{nuc}$ and annihilation distance $L_{ann}$ are $125b$ and $6b$. We obtain the Frank-Read properties from the formulas provided in [11]. The properties of precipitate are identical to the three-dimensional simulations presented in Section 2 with a diameter of $D=100$ nm, a spacing ratio of $L/D=5$, and a resistance ratio of $R=1$ (non-shearable precipitate). Performing three-dimensional dislocation dynamics simulations, we calculate the precipitate shear resistance $\tau'_c = 63.54$ MPa and the interaction time $t'_c = 0.6$ ns.

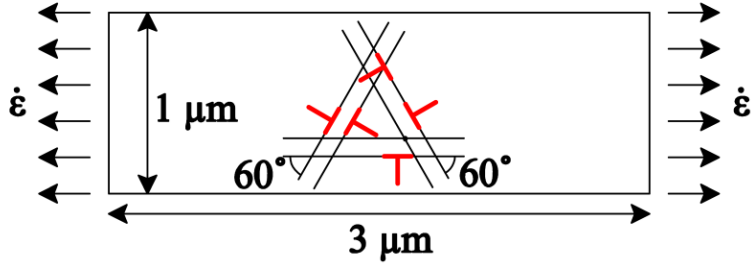

Fig. 10. Schematic representation of thin film model.

We carry out simulations for five scenarios: (1) a thin film without precipitates; (2) a thin film with a precipitate area fraction of 10% and neglected dislocation-precipitate interaction time $(t^* = 0)$; (3) a thin film with an identical distribution and properties of initial dislocations, Frank-Read sources, and precipitates to the second scenario while we model the dislocation-precipitate interaction time with the extended dislocation-precipitate interaction map presented in Section 3.3; scenarios (4) and (5) are similar to (2) and (3), respectively, while we perform simulations for a precipitates density of 20%. Figure 11 shows the distribution of initial dislocations, Frank-Read sources, and precipitates for all scenarios.

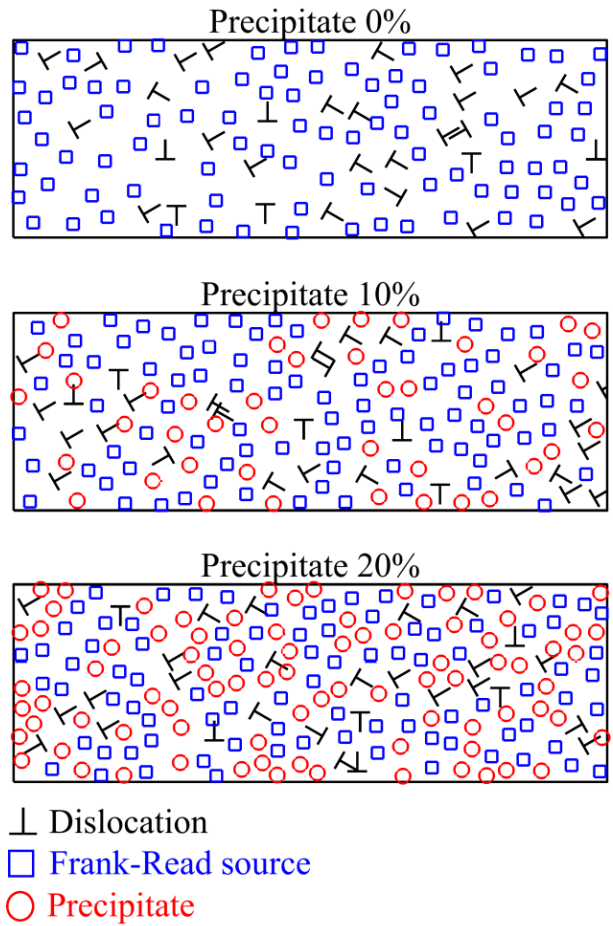

Fig. 11. Distribution of initial dislocations, Frank-Read sources, and precipitates.

Figure 12 shows the stress-strain curve for all scenarios. The results show that precipitates significantly increase the yield stress level. In addition, Fig. 12 enlightens the dual effect of the dislocation-precipitate interaction time. At early stages of plastic deformation, the thin film which contains precipitates with dislocation-precipitate interaction time model shows a higher level hardening. Specifically, for precipitate area fraction of 10%, the thin film with dislocation-precipitate interaction time model shows a higher level of hardening in a range of strain values between 0.21% and 0.56%. This range extends to a range of strain valsue between 0.22% and 0.87% as the area fraction of precipitates increases to 20%. The primary hardening stage results from the interaction between dislocations and precipitates. In former two-dimensional dislocation-precipitate interaction models, when a dislocation encounters a precipitate, the dislocation stops until the resolved shear sress from the dislocation reaches the precipitate resistance. However, with the presented dislocation-precipitate interaction time model, the dislocation stops for a period of time even after the resulting shear stress from the dislocation reaches the precipitate resistance. Therefore, precipitates with the dislocation-precipitate interaction time model result in more restricted dislocation movements leading to higher levels of hardening.

Conversely, the thin film with dislocation-precipitate interaction time model shows a lower hardening level at strain values higher than 0.56% and 0.87% for precipitate area fractions of 10% and 20%, respectively. This secondary hardening stage mainly results from dislocation pile-ups. Dislocation pile-ups grow slower when the dislocation-precipitate interaction time is modeled since precipitates with the dislocation-precipitate interaction time model leads to more restricted dislocation movements and delays the dislocation pile up formation. The stress field and dislocation distribution for a precipitate density of 20% with the dislocation-precipitate interaction model at an applied strain of 0.5% is presented in Figure 13. This figure shows the formation dislocation pile-ups, which is the secondary hardening mechanism.

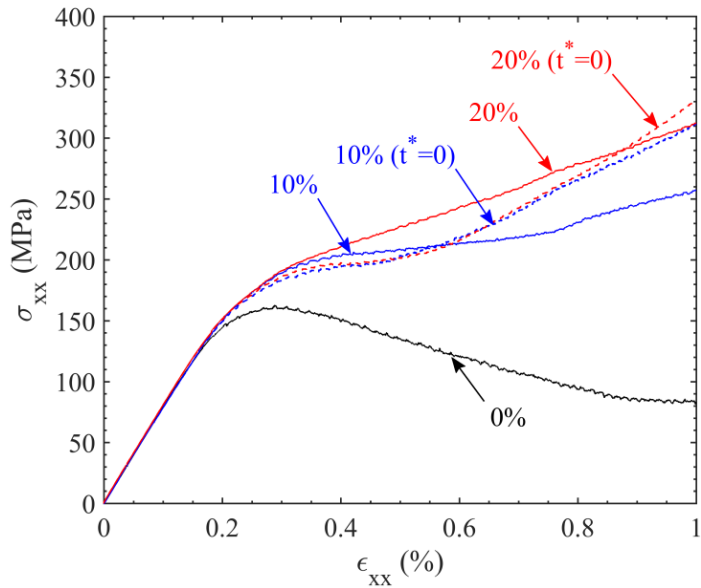

Fig. 12. Stress-strain curves for the copper thin film with various precipitates density and interaction model.

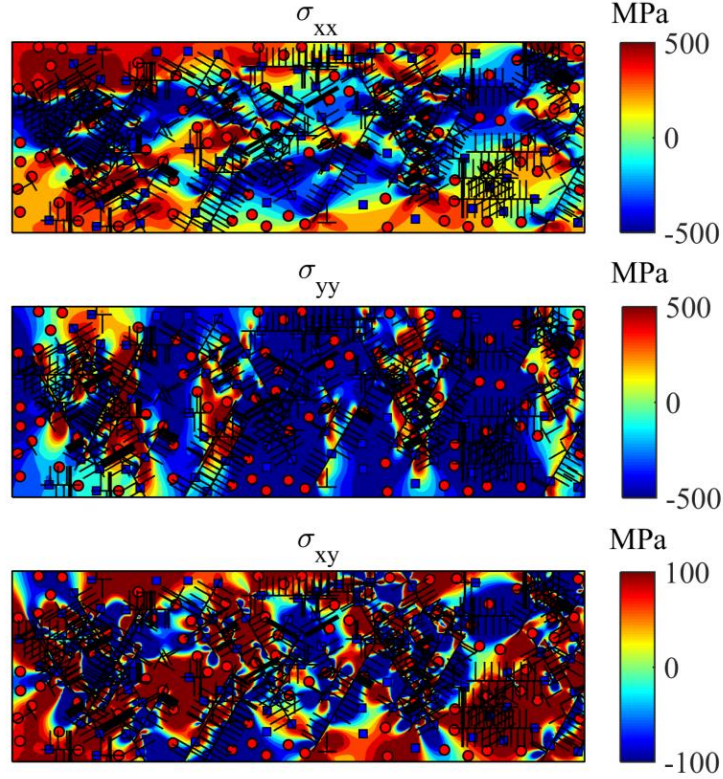

Fig. 13. Distribution of dislocations and stress field of the thin film at $\varepsilon = 0.5\%$ for a precipitate density of 20% and the dislocation-precipitate interaction time of $t^*$.

## 4. Conclusion

This study examined the interaction of an edge dislocation line with an array of collinear equally spaced precipitates using a modified line dislocation dynamics approach. We obtain thousands of data points from DD simulations that account for precipitate resistance and applied stress levels, and different precipitate spacing and introduce the dislocation-precipitate interaction time in a consistent structure with the nucleation time of the Frank-Read source. We developed a universal equation that holds for all simulations, called the "dislocation-interaction map." The outcome of this research is the most general equation for the interaction of an edge dislocation with an array of collinear equally spaced precipitates in terms of applied shear stress, precipitate resistance, and dislocation-precipitate interaction time. In addition, we extended the dislocation-interaction map for time-dependent applied shear stress scenarios. Previous studies on two-dimensional dislocation dynamics simulation (2D-DD) did not have the capability to model the dislocation-precipitate interaction time. We studied the mechanical behavior of a copper thin film at a high strain rate of $10^4 s^{-1}$ in a multi-scale framework based on the Finite Element Method (FEM) and the enriched 2D-DD with the dislocation-precipitate interaction map. Results enlightened the dual effect of the dislocation-precipitate interaction time on the hardening level. At the primary stage of plastic deformation, the dislocation-precipitate interaction time results in a higher level of hardening; however, the interaction time causes a lower level of hardening at the secondary stage of plastic deformation when hardening mainly results from dislocation pile-ups.

**References**

1. Foreman, A.J.E. and M.J. Makin, *Dislocation movement through random arrays of obstacles.* Philosophical magazine, 1966. **14**(131): p. 911-924.
2. Brown, L.M. and R.K. Ham, *Dislocation-particle interactions.* Strengthening Methods in Crystals, 1971: p. 9-135.
3. Ardell, A.J., *Precipitation hardening.* Metallurgical Transactions A, 1985. **16**(12): p. 2131-2165.
4. Amodeo, R.J. and N.M. Ghoniem, *Dislocation dynamics. I. A proposed methodology for deformation micromechanics.* Physical Review B, 1990. **41**(10): p. 6958.
5. Amodeo, R.J. and N.M. Ghoniem, *Dislocation dynamics. II. Applications to the formation of persistent slip bands, planar arrays, and dislocation cells.* Physical Review B, 1990. **41**(10): p. 6968.
6. Gulluoglu, A.N. and C.S. Hartley, *Simulation of dislocation microstructures in two dimensions. I. Relaxed structures.* Modelling and Simulation in Materials Science and Engineering, 1992. **1**(1): p. 17.
7. Gulluoglu, A.N. and C.S. Hartley, *Simulation of dislocation microstructures in two dimensions. II. Dynamic and relaxed structures.* Modelling and Simulation in Materials Science and Engineering, 1993. **1**(4): p. 383.
8. Kondori, B., A. Needleman, and A. Amine Benzerga, *Discrete dislocation simulations of compression of tapered micropillars.* Journal of the Mechanics and Physics of Solids, 2017. **101**: p. 223-234.
9. Huang, M., L. Zhao, and J. Tong, *Discrete dislocation dynamics modelling of mechanical deformation of nickel-based single crystal superalloys.* International Journal of Plasticity, 2012. **28**(1): p. 141-158.
10. Shin, C.S., et al., *Dislocation–impenetrable precipitate interaction: a three-dimensional discrete dislocation dynamics analysis.* Philosophical Magazine, 2003. **83**(31-34): p. 3691-3704.
11. Van der Giessen, E. and A. Needleman, *Discrete dislocation plasticity: a simple planar model.* Modelling and Simulation in Materials Science and Engineering, 1995. **3**(5): p. 689.
12. Zbib, H.M. and T. Diaz de la Rubia, *A multiscale model of plasticity.* International Journal of Plasticity, 2002. **18**(9): p. 1133-1163.
13. Takahashi, A. and N.M. Ghoniem, *A computational method for dislocation–precipitate interaction.* Journal of the Mechanics and Physics of Solids, 2008. **56**(4): p. 1534-1553.
14. Keyhani, A., et al., *XFEM–dislocation dynamics multi-scale modeling of plasticity and fracture.* Computational Materials Science, 2015. **104**: p. 98-107.
15. Yashiro, K., et al., *Discrete dislocation dynamics simulation of cutting of γ' precipitate and interfacial dislocation network in Ni-based superalloys.* International Journal of Plasticity, 2006. **22**(4): p. 713-723.
16. Gao, S., et al., *Influence of misfit stresses on dislocation glide in single crystal superalloys: A three-dimensional discrete dislocation dynamics study.* Journal of the Mechanics and Physics of Solids, 2015. **76**: p. 276-290.
17. Xu, S., et al., *Edge dislocations bowing out from a row of collinear obstacles in Al.* Scripta Materialia, 2016. **123**: p. 135-139.

18. Osetsky†, Y.N., D.J. Bacon, and V. Mohles, *Atomic modelling of strengthening mechanisms due to voids and copper precipitates in α-iron.* Philosophical Magazine, 2003. **83**(31-34): p. 3623-3641.

19. Lehtinen, A., et al., *Multiscale modeling of dislocation-precipitate interactions in Fe: From molecular dynamics to discrete dislocations.* Physical Review E, 2016. **93**(1): p. 013309.

20. Naveen Kumar, N., et al., *Modeling of radiation hardening in ferritic/martensitic steel using multi-scale approach.* Computational Materials Science, 2012. **53**(1): p. 258-267.

21. Diaz de la Rubia, T., et al., *Multiscale modelling of plastic flow localization in irradiated materials.* Nature, 2000. **406**(6798): p. 871-874.

22. de Koning, M., W. Cai, and V.V. Bulatov, *Anomalous Dislocation Multiplication in FCC Metals.* Physical Review Letters, 2003. **91**(2): p. 025503.

23. Argon, A., *Strengthening mechanisms in crystal plasticity*. 2008: Oxford University Press on Demand.

24. Khraishi, T.A., L. Yan, and Y.L. Shen, *Dynamic simulations of the interaction between dislocations and dilute particle concentrations in metal–matrix composites (MMCs).* International Journal of Plasticity, 2004. **20**(6): p. 1039-1057.

25. Mohles, V., *Simulations of dislocation glide in overaged precipitation-hardened crystals.* Philosophical Magazine A, 2001. **81**(4): p. 971-990.

26. Mohles, V., *The critical resolved shear stress of single crystals with long-range ordered precipitates calculated by dislocation dynamics simulations.* Materials Science and Engineering: A, 2004. **365**(1): p. 144-150.

27. Mohles, V., D. Rönnpagel, and E. Nembach, *Simulation of dislocation glide in precipitation hardened materials.* Computational Materials Science, 1999. **16**(1–4): p. 144-150.

28. Xiang, Y., L.T. Cheng, and D.J. Srolovitz, *A level set method for dislocation dynamics.* Acta materialia, 2003. **51**(18): p. 5499-5518.

29. Xiang, Y., D.J. Srolovitz, and L.T. Cheng, *Level set simulations of dislocation-particle bypass mechanisms.* Acta materialia, 2004. **52**(7): p. 1745-1760.

30. Monnet, G., *Investigation of precipitation hardening by dislocation dynamics simulations.* Philosophical Magazine, 2006. **86**(36): p. 5927-5941.

31. Mohles, V., *Computer simulations of particle strengthening: lattice mismatch strengthening.* Materials Science and Engineering: A, 2001. **319–321**: p. 201-205.

32. Mohles, V., *Computer simulations of particle strengthening: the effects of dislocation dissociation on lattice mismatch strengthening.* Materials Science and Engineering: A, 2001. **319–321**: p. 206-210.

33. Mohles, V., *Computer simulations of the glide of dissociated dislocations in lattice mismatch strengthened materials.* Materials Science and Engineering: A, 2002. **324**(1–2): p. 190-195.

34. Mohles, V., *Superposition of dispersion strengthening and size-mismatch strengthening: Computer simulations.* Philosophical Magazine Letters, 2003. **83**(1): p. 9-19.

35. Duesbery, M.S. and K. Ssadananda, *The interaction of dislocations with coherent inclusions I. Perfect edge and screw dislocations.* Philosophical Magazine A, 1991. **63**(3): p. 535-558.

36. Keyhani, A., R. Roumina, and S. Mohammadi, *An efficient computational technique for modeling dislocation–precipitate interactions within dislocation dynamics.* Computational Materials Science, 2016. **122**: p. 281-287.

37. Ryu, S., K. Kang, and W. Cai, *Entropic effect on the rate of dislocation nucleation.* Proceedings of the National Academy of Sciences, 2011. **108**(13): p. 5174-5178.

38. Kocks, U.F., *Kinetics of solution hardening.* Metallurgical Transactions A, 1985. **16**(12): p. 2109-2129.

39. Xu, Z. and R.C. Picu, *Thermally activated motion of dislocations in fields of obstacles: The effect of obstacle distribution.* Physical Review B, 2007. **76**(9): p. 094112.

40. Picu, R.C., R. Li, and Z. Xu, *Strain rate sensitivity of thermally activated dislocation motion across fields of obstacles of different kind.* Materials Science and Engineering: A, 2009. **502**(1–2): p. 164-171.

41. Torabi, M., A. Keyhani, and G.P. Peterson, *A comprehensive investigation of natural convection inside a partially differentially heated cavity with a thin fin using two-set lattice Boltzmann distribution functions.* International Journal of Heat and Mass Transfer, 2017. **115**(Part A): p. 264-277.

42. Keyhani, A. and R. Roumina, *Dislocation-precipitate interaction map.* Computational Materials Science, 2018. **141**: p. 153-161.

43. Bulatov, V.V. and W. Cai, *Computer simulations of dislocations*. Vol. 3. 2006: Oxford University Press.

44. Humphreys, F.J. and P.B. Hirsch. *The deformation of single crystals of copper and copper-zinc alloys containing alumina particles. ii. microstructure and dislocation-particle interactions*. in *Proceedings of the Royal Society of London A: Mathematical, Physical and Engineering Sciences*. 1970. The Royal Society.

45. Hirsch, P.B., *The interpretation of the slip pattern in terms of dislocation movements.* Journal of the Institute of Metals, 1957. **86**: p. 7.

46. Bacon, D., U. Kocks, and R. Scattergood, *The effect of dislocation self-interaction on the Orowan stress.* Philosophical Magazine, 1973. **28**(6): p. 1241-1263.

47. Martínez, E., et al., *Atomistically informed dislocation dynamics in fcc crystals.* Journal of the Mechanics and Physics of Solids, 2008. **56**(3): p. 869-895.

48. Foreman, A.J.E., *The bowing of a dislocation segment.* Philosophical magazine, 1967. **15**(137): p. 1011-1021.

49. Benzerga, A.A., *An analysis of exhaustion hardening in micron-scale plasticity.* International Journal of Plasticity, 2008. **24**(7): p. 1128-1157.

50. Benzerga, A.A., et al., *Incorporating three-dimensional mechanisms into two-dimensional dislocation dynamics.* Modelling and Simulation in Materials Science and Engineering, 2004. **12**(1): p. 159.

51. Keralavarma, S.M. and W.A. Curtin, *Strain hardening in 2D discrete dislocation dynamics simulations: A new '2.5D' algorithm.* Journal of the Mechanics and Physics of Solids, 2016. **95**: p. 132-146.

52. Davoudi, K.M., L. Nicola, and J.J. Vlassak, *Dislocation climb in two-dimensional discrete dislocation dynamics.* Journal of Applied Physics, 2012. **111**(10): p. 103522.

53. Rohde, R.W. and C.H. Pitt, *Dislocation Velocities in Nickel Single Crystals.* Journal of Applied Physics, 1967. **38**(2): p. 876-879.

54. Stein, D.F. and J.R.L. Jr., *Mobility of Edge Dislocations in Silicon-Iron Crystals.* Journal of Applied Physics, 1960. **31**(2): p. 362-369.

55. Johnston, W.G. and J.J. Gilman, *Dislocation Velocities, Dislocation Densities, and Plastic Flow in Lithium Fluoride Crystals.* Journal of Applied Physics, 1959. **30**(2): p. 129-144.

56. Greenman, W.F., T.V. Jr., and D.S. Wood, *Dislocation Mobility in Copper.* Journal of Applied Physics, 1967. **38**(9): p. 3595-3603.